# Quantum capacitance mediated carbon nanotube optomechanics

Stefan Blien[1], Patrick Steger[1], Niklas Hüttner[1], Richard Graaf[1] & Andreas K. Hüttel[1,2]✉

Cavity optomechanics allows the characterization of a vibration mode, its cooling and quantum manipulation using electromagnetic fields. Regarding nanomechanical as well as electronic properties, single wall carbon nanotubes are a prototypical experimental system. At cryogenic temperatures, as high quality factor vibrational resonators, they display strong interaction between motion and single-electron tunneling. Here, we demonstrate large optomechanical coupling of a suspended carbon nanotube quantum dot and a microwave cavity, amplified by several orders of magnitude via the nonlinearity of Coulomb blockade. From an optomechanically induced transparency (OMIT) experiment, we obtain a single photon coupling of up to $g_0 = 2\pi \cdot 95$ Hz. This indicates that normal mode splitting and full optomechanical control of the carbon nanotube vibration in the quantum limit is reachable in the near future. Mechanical manipulation and characterization via the microwave field can be complemented by the manifold physics of quantum-confined single electron devices.

[1] Institute for Experimental and Applied Physics, University of Regensburg, 93040 Regensburg, Germany. [2]Present address: Department of Applied Physics, Aalto University, Puumiehenkuja 2, 02150 Espoo, Finland. ✉email: andreas.huettel@ur.de

The technically challenging integration of suspended single-wall cabon nanotubes into complex qantum devices has recently made significant advances[1–6], as has also the integration of nanotube quantum dots into coplanar microwave cavities[7–9]. Both regarding their nanomechanical[10,11] as well as their electronic properties[12,13], carbon nanotubes are a prototypical experimental system. However, small vibrational deflection and length have made their optomechanical coupling to microwave fields[14] so far impossible.

In this work, we demonstrate large optomechanical coupling of a suspended carbon nanotube quantum dot and a microwave cavity. The nanotube is deposited onto source and drain electrodes close to the coplanar waveguide cavity; a finger-like extension of the cavity center conductor, passing below the suspended nanotube, serves as capacitively coupling gate. We find that the optomechanical coupling of the transversal nanotube vibration and the cavity mode is amplified by several orders of magnitude via the inherent nonlinearity of Coulomb blockade. With this, full optomechanical control of the carbon nanotube vibration in the quantum limit[15] is reachable in the near future. A unique experimental system becomes accessible, where the nanomechanically active part directly incorporates a quantum-confined electron system[16].

## Results

**Device precharacterization.** Our device, depicted in Fig. 1a, combines a half-wavelength coplanar microwave cavity with a suspended carbon nanotube quantum dot. Near the coupling capacitor, the center conductor of the niobium-based cavity is connected to a thin gate electrode, buried between source and drain contacts of the carbon nanotube, see the sketch of Fig. 1b. At the cavity center, i.e., the location of the voltage node of its fundamental mode, a bias connection allows additional application of a dc voltage $V_g$ to the gate. The device is mounted at the base temperature stage ($T \simeq 10$ mK) of a dilution refrigerator; for details see Supplementary Note 4 and Supplementary Fig. 4.

At cryogenic temperatures, electronic transport through the carbon nanotube is dominated by Coulomb blockade, with the typical behavior of a small band gap nanotube[12]. Near the electronic band gap, sharp Coulomb oscillations of conductance can be resolved; measurements are shown in Fig. 1c and Supplementary Fig. 3. A well-known method to detect the transversal vibration resonance of a suspended nanotube quantum dot is to apply a rf signal and measure the time-averaged dc current[17–19]. On resonance, the oscillating geometric capacitance, effectively broadening the Coulomb oscillations, leads to an easily recognizable change in current. This was used to identify the transversal vibration resonances of the device; Fig. 1d plots the resonance frequencies over a wide gate voltage range. Two coupled vibration modes are observed (see also Supplementary Note 5), one of which clearly displays electrostatic softening[20,21]. At low gate voltages, $|V_g| \leq 1.2$ V, where subsequent experiments are carried out, the resonance which we will utilize in the following is at $\omega_m \simeq 2\pi \cdot 502.5$ MHz, with typical quality factors around or exceeding $Q_m \sim 10^4$ observed in time-averaged dc current detection[17].

The combined suspended nanotube—cavity device forms a dispersively coupled optomechanical system[14]. The cavity has a resonance frequency of $\omega_c = 2\pi \cdot 5.74$ GHz with a decay rate of $\kappa_c = 2\pi \cdot 11.6$ MHz, dominated by internal losses. Nevertheless, due to the large mechanical resonance frequency $\omega_m$ of the carbon nanotube, the coupled system is far in the resolved sideband regime $\omega_m \gg \kappa_c$, the most promising parameter region for a large number of optomechanical protocols including ground state cooling and quantum control.

**Optomechanically induced transparency (OMIT).** To probe for optomechanical coupling, we perform an OMIT type experiment[22], cf. Fig. 2a, b: a strong, red-detuned drive field ($\omega_d \simeq \omega_c - \omega_m$) pumps the microwave cavity; the transmission of a weak, superimposed probe signal $\omega_p$ near $\omega_c$ is detected. A distinct, sharp OMIT absorption feature within the transmission resonance of the cavity becomes visible in the measurements of Fig. 2c–e. It occurs due to destructive interference of the probe field with optomechanically upconverted photons of the drive field, when the two-photon resonance condition $\omega_p - \omega_d = \omega_m$ is fulfilled[22], and shifts in frequency as expected when $\omega_d$ is detuned from the precise red sideband condition, see Fig. 2d, e. Fitting the OMIT feature allows to extract the optomechanical coupling parameter $g = \sqrt{n_c}(\partial \omega_c/\partial x)x_{zpf}$, describing the cavity detuning per displacement of the mechanical harmonic oscillator[14,22], see Supplementary Note 9 for details. Surprisingly, from Fig. 2c, one obtains a single-photon coupling on the order of $g_0 = g/\sqrt{n_c} \sim 2\pi \cdot 100$ Hz.

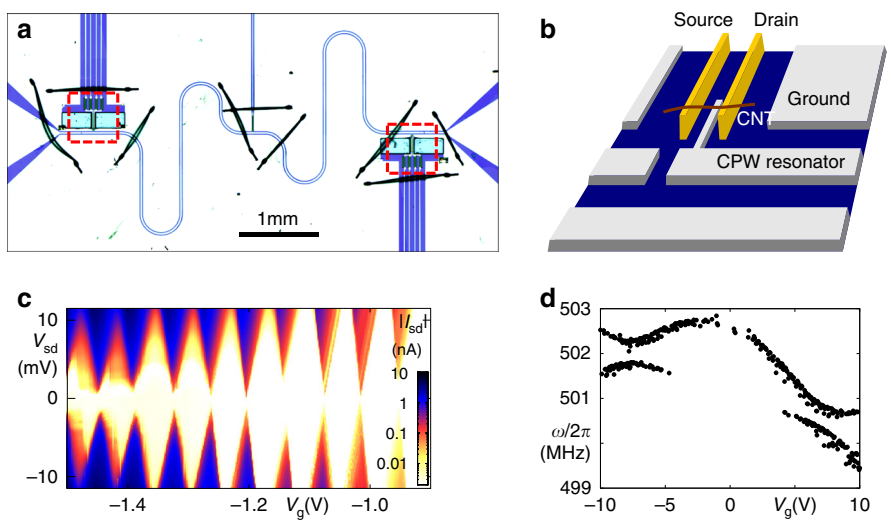

**Fig. 1 Integrating a suspended carbon nanotube into a microwave cavity. a** Optical micrograph showing a niobium-based $\lambda/2$ coplanar waveguide cavity for transmission measurement, with carbon nanotube deposition areas and dc contact structures (see the red dashed squares) near the coupling capacitors. For fabrication redundancy, two deposition areas exist on the device, but only one is used here. Bond wires visible as dark lines connect different segments of the ground plane to avoid spurious resonances. **b** Simplified sketch of the nanotube deposition area, including source and drain electrodes, a carbon nanotube deposited on them, and the buried gate connected to the cavity center conductor. **c** dc transport characterization of the carbon nanotube at $T_{base} \simeq 10$ mK. The plot of the absolute value of currrent $|I|$ as function of gate voltage $V_g$ and bias voltage $V_{sd}$ displays the typical diamond-shaped Coulomb blockade regions of suppressed conductance[12,40,41]. **d** Using rf excitation with an antenna and dc measurement[17,18], two transversal vibration modes can be traced across a large gate voltage range; the figure plots the detected resonance frequencies. The corresponding raw data as well as a fit can be found in Supplementary Figs. 6 and 7.

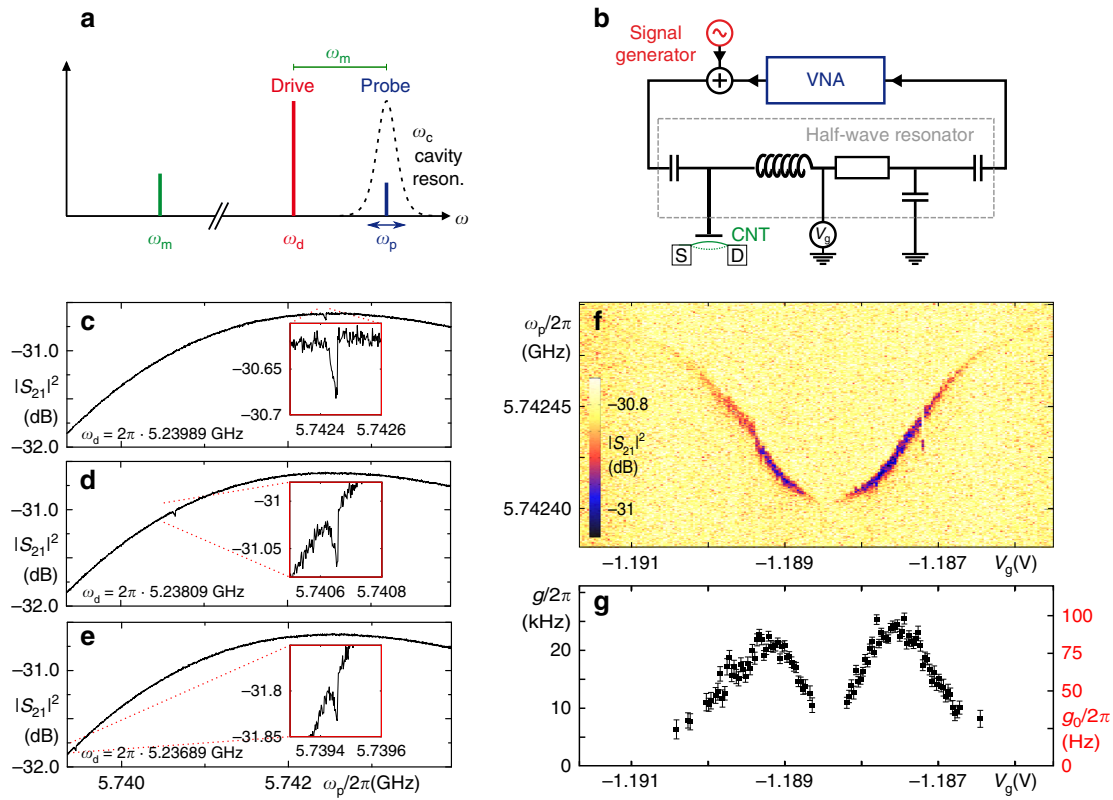

**Fig. 2 Optomechanically induced transparency (OMIT) in the Coulomb blockade regime. a** Frequency scheme and **b** detection setup of an OMIT measurement. A strong drive signal at $\omega_d = \omega_c - \omega_m$ pumps the microwave cavity; the cavity transmission near the cavity resonance $\omega_c$ is characterized using a superimposed weak probe signal $\omega_p$ from a vector network analyzer (VNA). Device parameters are: $\omega_c \simeq 2\pi \cdot 5.74$ GHz, $\kappa_c = 2\pi \cdot 11.6$ MHz, $\omega_m \simeq 2\pi \cdot 502.5$ MHz. **c–e** Probe signal power transmission $|S_{21}(\omega_p)|^2$ for three different choices of cavity drive frequency $\omega_d$, at $\omega_d = \omega_c - \omega_m$ (**c**) and slightly detuned (**d**, **e**). The gate voltage $V_g = -1.1855$ V is fixed on the flank of a sharp Coulomb oscillation of conductance; $V_{sd} = 0$. **f** Probe signal transmission as in **c–e**, now for a fixed cavity drive frequency $\omega_d = 2\pi \cdot 5.23989$ GHz and varied gate voltage $V_g$ across a Coulomb oscillation. The depth of the OMIT feature allows the evaluation of the optomechanical coupling $g(V_g)$ at each gate voltage value. **g** Optomechanical coupling $g(V_g)$ (left axis) and corresponding single photon coupling $g_0(V_g) = g(V_g)/\sqrt{n_c}$ (right axis), extracted from the data of **f**; $n_c = 67{,}500$. Error bars indicate the standard error of the fit result.

Such a value of $g_0$ strongly exceeds expectations from the device geometry[23]. For a mechanical oscillator dispersively coupled to a coplanar waveguide resonator, the coupling is given by

$$g_0 = \frac{\omega_c}{2C_c}\frac{\partial C_c}{\partial x}x_{zpf}, \qquad (1)$$

where $C_c$ is the total capacitance of the cavity, $x$ is the mechanical displacement, and $x_{zpf}$ the mechanical zero-point fluctuation length scale. Assuming a metallic wire over a metallic plane and inserting device parameters[23], the coupling calculated from the change in geometric gate capacitance $C_g(x)$ becomes $\partial C_g/\partial x \sim 10^{-12}$ F m$^{-1}$. This leads to $g_0^* = 2\pi \cdot 2.9$ mHz, more than four orders of magnitude smaller than the measured $g_0$. To explain this discrepancy, we need to focus on the properties of the carbon nanotube as a quantum dot, with a strongly varying quantum capacitance $C_{CNT}(x)$ as the displacement-dependent component of $C_c$ dominating $g_0$.

Figure 2f depicts OMIT measurements for similar parameters as in Fig. 2c–e, however, we now keep the drive frequency $\omega_d$ constant and vary the gate voltage $V_g$ across a Coulomb oscillation of conductance. The mechanical resonance frequency $\omega_m$ shifts to lower frequencies in the vicinity of the charge degeneracy point. This electrostatic softening is a well-known characteristic of suspended carbon nanotube quantum dots[18,24]. More interestingly, the resulting gate-dependent coupling $g(V_g)$ (along with $g_0(V_g)$) is plotted in Fig. 2g. It is maximal at the edges of the finite conductance peak, whereas at its center and on the outer edges, the coupling vanishes; the enhancement of $g_0$ is intrinsically related to Coulomb blockade.

**Mechanism of enhanced coupling.** Figure 3 explores the nature of this enhanced coupling mechanism. We treat the nanotube as a single quantum dot; see Supplementary Note 3 for a discussion of the validity of this assumption. Further, we assume a full separation of time scales $\omega_m \ll \omega_c \ll \Gamma$, where $\Gamma$ describes the tunnel rates of the quantum dot. We can then introduce the quantum capacitance[25,26]

$$C_{CNT} = e\frac{C_g}{C_{dot}}\frac{\partial \langle N \rangle}{\partial V_g}, \qquad (2)$$

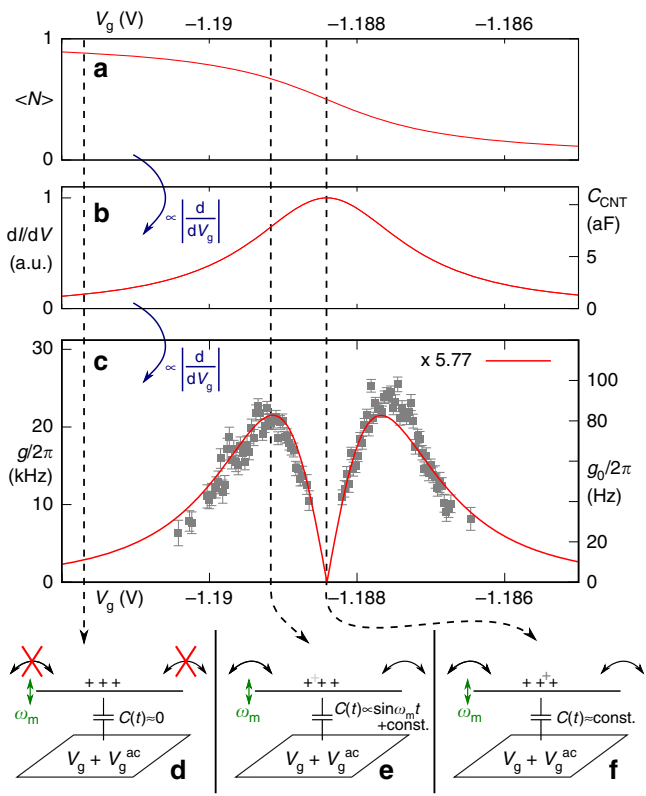

**Fig. 3 Coulomb blockade enhanced optomechanical coupling mechanism.** Solid lines correspond to the model of a Lorentz-broadened quantum dot level at $k_B T \ll \Gamma$. The Coulomb oscillation center $V_g^* = -1.18841$ V, the line width $\hbar\Gamma = 0.673$ meV, and a scaling prefactor $a = 5.77$ (see text) have been obtained by fitting to the OMIT data $g(V_g)$. **a** Time-averaged charge occupation $\langle N \rangle(V_g)$ of the quantum dot (note that we are in the hole conduction regime). **b** Conductance $dI/dV_{sd}(V_g)$ (left axis) and quantum capacitance $C_{CNT}$ (right axis), cf. Supplementary Fig. 12. **c** Coulomb-blockade enhanced optomechanical coupling $g(V_g)$ (left axis) and single photon coupling $g_0(V_g)$ (right axis). The data points are identical to Fig. 2g; the calculation result has been scaled with 5.77 to fit the data. **d–f** Schemata for the situations corresponding to the dashed lines in **a–c**, see the text.

where $\langle N \rangle(V_g)$ is the number of charge carriers (here holes) on the quantum dot averaged over the tunneling events, and $C_{dot}$ is the total quantum dot capacitance; see Supplementary Note 12 for a derivation. In a quantum dot, each Coulomb oscillation corresponds to the addition of one electron or hole. The charge occupation $\langle N \rangle(V_g)$ resembles a step function, with the sharpness of the step given for zero bias voltage by lifetime and temperature broadening. This is plotted in Fig. 3a, for the limit of $k_B T \ll \Gamma$. The quantum capacitance $C_{CNT}(V_g)$ becomes a Lorentzian, as plotted in Fig. 3b.

Any motion $\delta x$ modulates the geometric capacitance $C_g(x)$. It thus shifts the position of the Coulomb oscillations in gate voltage, acting equivalent to an effective modulation of the gate voltage $\delta V_g$. With this, the optomechanical coupling $g$, scaling with $|\partial C_{CNT}/\partial x|$, becomes proportional to the derivative $\partial C_{CNT}/\partial V_g$ and thus the second derivative of $\langle N \rangle(V_g)$, as is illustrated in Fig. 3c. The functional dependence has been fitted to the data points of Fig. 2g, here again shown in the background.

The three key situations depending on the gate voltage are sketched in Fig. 3d–f: away from the conductance peak, the charge on the nanotube is constant, and only geometric capacitances change, see Fig. 3d. On the flank of the conductance resonance, a small change $\delta x$ ($\propto \delta C_g$) strongly modulates $C_{CNT}$, see Fig. 3e. At the center of the conductance resonance, the charge adapts to $x$, but the derivative $\partial C_{CNT}/\partial V_g$ and with it $g \propto |\partial C_{CNT}/\partial x|$ is approximately zero.

The detailed derivation and the full expressions and values for Fig. 3 can be found in the Supplementary Information, Supplementary Note 12, and Supplementary Table 1. The parameter entering the optomechanical coupling in Eq. (1), the derivative of the quantum capacitance $\partial C_{CNT}/\partial x$, is found to be

$$\frac{\partial C_{CNT}}{\partial x} = \eta \frac{\partial C_g}{\partial x} = e \frac{\partial^2 \langle N \rangle}{\partial V_g^2} \frac{V_g}{C_{dot}} \frac{\partial C_g}{\partial x}, \quad (3)$$

indicating that for significant optomechanical coupling a sharp Coulomb oscillation (i.e., low temperature and low intrinsic line width $\Gamma$, leading to large values of $\partial^2 \langle N \rangle/\partial V_g^2$) and a large $V_g$ are required. From device data, we obtain an amplification factor $\eta \sim 10^4$. The experimental gate voltage dependence $g_0(V_g)$ is qualitatively reproduced very well. To obtain the quantitative agreement of Fig. 3c, we have introduced an additional scaling prefactor as free fit parameter, resulting in $g_0^{exp}/g_0^{th} = 5.77$. Given the uncertainties of input parameters, this is a good agreement; see Supplementary Note 15 for a discussion of error sources.

## Discussion

In literature, many approaches have been pursued to enhance optomechanical coupling[26–35]. Resonant coupling, with $\omega_m = \omega_c$, has been demonstrated successfully for a carbon nanotube quantum dot[26], but does not provide access to the wide set of experimental protocols developed for the usual case of dispersive coupling and the "good cavity limit" $\omega_m \gg \kappa_c$. The mechanism presented here is most closely related to those where a superconducting charge qubit was coherently introduced between mechanical resonator and cavity[27]. However, the impact of single electron tunneling and shot noise on the optomechanical system shall require careful analysis.

Given the sizeable coupling in the good cavity limit $\kappa_c \ll \omega_m$, many experimental techniques for future experiments are at hand. First steps are demonstrated in Fig. 4 in a two-tone spectroscopy experiment: a mechanical drive signal $\omega_a$ is applied simultaneously to a cavity pump signal at $\omega_d = \omega_c - \omega_a$; the plotted cavity output power at $\omega_c$ clearly shows the optmechanical upconversion (anti-Stokes scattering) at mechanical resonance $\omega_a = \omega_m$. In Fig. 4a, the dc bias across the nanotube is set to zero, and the antenna drive kept at a minimum. In Fig. 4b, both antenna drive and bias voltage have been increased. A background signal independent of device parameters emerges; at the same time, the upconverted signal displays a phase shift and destructive interference with the background for parts of the gate voltage range, meriting further measurements and analysis.

Future improvements of the optomechanical coupling via drive power and device geometry and of the detection sensitivity via the output amplifier chain shall allow detection of the thermal motion of the carbon nanotube and subsequently motion amplitude calibration.

The observation of strong optomechanical coupling and the corresponding normal mode splitting requires a coupling $g$ exceeding both mechanichal linewidth $\kappa_m$ and cavity line width $\kappa_c$. Clean carbon nanotubes have reached mechanical quality factors up to[36] $Q_m \sim 10^6$, allowing for two orders of magnitude improvement and a line width of $\kappa_m \sim 2\pi \cdot 500$ Hz. Regarding microwave resonators we have reached up to $Q_c = 10^5$ in our setup so far, corresponding to $\kappa_c = 2\pi \cdot 57$ kHz. This means that strong coupling should be reachable already at moderate increase of our so far rather low cavity photon number $n_c$.

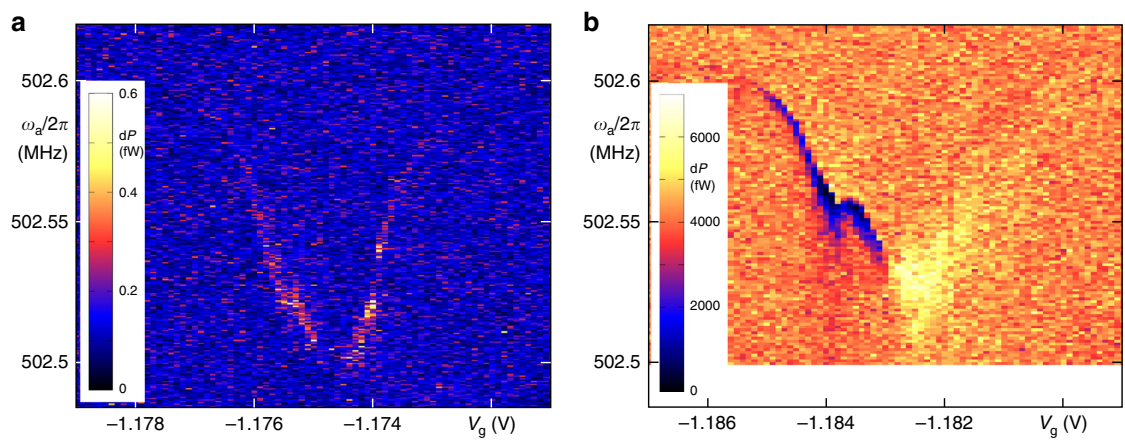

**Fig. 4 Two-tone spectroscopy.** Via an antenna, the carbon nanotube is driven at $\omega_a$ close its mechanical eigenfrequency; the microwave cavity is simultaneously pumped at $\omega_d = \omega_c - \omega_a$. The plots show the power output of the cavity at the upconverted frequency $\omega_c$, with the nanotube acting as nonlinear element. Drive power $P_d = 20$ dBm ($n_c \simeq 2.1 \times 10^4$), measurement bandwidth 5 Hz. **a** Antenna generator power $P_a = -55$ dBm, bias $V_{sd} = 0$; **b** antenna generator power $P_a = -30$ dBm, bias $V_{sd} = 0.5$ mV.

Regarding the cooperativity $C = 0.0042$ of our experiment (cf. Supplementary Table 1), already an improvement of the nanotube $Q_m$ by a factor 100 brings it into the same order of magnitude as the thermal mode occupation $n_m = 0.4$, with significant further and independent room for improvement via the cavity photon number $n_c$.

With this, a wide range of physical phenomena becomes experimentally accessible, ranging from side-band cooling of the vibration mode and potentially its quantum control[37] all the way to real-time observation of its interaction with single electron tunneling phenomena[38].

### Data availability
The datasets generated during and/or analyzed during this study are available from the corresponding author on reasonable request.

### Acknowledgements
The authors acknowledge funding by the Deutsche Forschungsgemeinschaft via Emmy Noether grant Hu 1808/1, SFB 631, SFB 689, SFB 1277, and GRK 1570. We would like to thank G. Rastelli, F. Marquardt, E. A. Laird, Y. M. Blanter, and D. Weiss for insightful discussions, O. Vavra for experimental help, and Ch. Strunk and D. Weiss for the use of the experimental facilities. The data have been recorded using Lab::Measurement[39].

### Author contributions
A.K.H. and S.B. conceived and designed the experiment. P.S. and R.G. developed and performed the nanotube growth and transfer; N.H. and S.B. developed and fabricated the coplanar waveguide device. The low temperature measurements were performed jointly by all authors. Data evaluation and writing of the paper was done jointly by S.B., N.H., and A.K.H. The project was supervised by A.K.H.

### Competing interests
The authors declare no competing interests.

### Additional information
**Supplementary information** is available for this paper at https://doi.org/10.1038/s41467-020-15433-3.

**Correspondence** and requests for materials should be addressed to A.K.Hüt.

**Peer review information** *Nature Communications* thanks Joel Moser and the other anonymous reviewer(s) for their contribution to the peer review of this work.

**Reprints and permission information** is available at http://www.nature.com/reprints

**Publisher's note** Springer Nature remains neutral with regard to jurisdictional claims in published maps and institutional affiliations.